# A Deep Multi-Level Attentive network for Multimodal Sentiment Analysis

Ashima Yadav[1], Dinesh Kumar Vishwakarma[2]
Biometric Research Laboratory, Department of Information Technology, Delhi Technological University, Delhi, India.
E-mail: [1]ashimayadavdtu@gmail.com, [2]dinesh@dtu.ac.in

**Abstract**— Multimodal sentiment analysis has attracted increasing attention with broad application prospects. The existing methods focuses on single modality, which fails to capture the social media content for multiple modalities. Moreover, in multimodal learning, most of the works have focused on simply combining the two modalities, without exploring the complicated correlations between them. This resulted in dissatisfying performance for multimodal sentiment classification. Motivated by the status quo, we propose a Deep Multi-Level Attentive network (DMLANet), which exploits the correlation between image and text modalities to improve multimodal learning. Specifically, we generate the bi-attentive visual map along the spatial and channel dimensions to magnify CNN's representation power. Then we model the correlation between the image regions and semantics of the word by extracting the textual features related to the bi-attentive visual features by applying semantic attention. Finally, self-attention is employed to automatically fetch the sentiment-rich multimodal features for the classification. We conduct extensive evaluations on four real-world datasets, namely, MVSA-Single, MVSA-Multiple, Flickr, and Getty Images, which verifies the superiority of our method.

## 1 INTRODUCTION

THE rapid popularity of social media has contributed to the enormous amount of multimedia data containing visual contents and textual descriptions. People are continuously expressing their emotions and opinions on social networking sites like Twitter, Flickr, etc. [1]. Sentiment analysis [2] of multimodal data is crucial for understanding the attitude and behavior of people in many real-world applications like healthcare [3], politics [4], cinematography [5], and business analysis [6]. Hence, automatically detecting the sentiments from visual and textual contents has emerged as a significant research problem.

Previous multimodal sentiment analysis works have concentrated on early fusion-based techniques that combine the different features from multiple sources and feed them into the classifier [7]. Some have predicted the sentiments from multiple sources and then aggregated the results to get the multimodal sentiment label [8]. This is known as late fusion. The major drawback of these works is that they fail to capture the complex correlation between the modalities. Similarly, some works have employed intermediate fusion, which combines the modalities in the intermediate layers of the network [9]. However, this requires a careful design approach and may not perform well if a portion of multimodal content is incomplete.

Although significant efforts have been made in this area, predicting the multimodal sentiments remains an open problem due to the following reasons. Firstly, each modality has its individual characteristics and is expressed differently by the human cognitive system [10]. Hence, it isn't easy to deal with such heterogeneous content for multimodal analysis. Therefore, the correlations between the image and text descriptions needs to be captured effectively to bridge this gap. Secondly, some of the existing works have focused on the attention mechanism to capture the crucial image regions and sentimental words for generating efficient multimodal features. However, these approaches only focus on region-based attention and do not utilize channel information for developing visual features. This is important because the channel-based attention helps to identify the crucial patterns in the given image. Thirdly, most works fail to utilize image-level features to highlight the sentimental words, which could enhance the performance for multimodal sentiment classification.

Hence, to tackle the above challenges, we propose a Deep Multi-level Attentive network (DMLANet) by introducing the channel-level and region-level attention to generate a bi-attentive visual feature map that enhances the representation power of the CNN. After that, we utilize the semantic attention to extract the attended textual features corresponding to the bi-attentive visual features. Next, we concatenate the attended text features with bi-attentive visual features to obtain the multimodal features, which are passed into the self-attention network, weighing all the multimodal features and extracting only the significant sentiment-rich features for the classification. Thus, our model exploits the fine-grained correlations between the image and text descriptions leading to effective classification results. The significant contribution can be summarized as follows:

- A Deep Multi-level Attentive network (DMLANet) is developed, which generates the discriminative features from the visual and textual descriptions by introducing attention at multiple levels. We obtain the bi-attentive visual features by exploiting the channel attention and spatial attention in the visual data.
- To capture the complicated correlation between the images and text, we develop a joint multimodal learning strategy that focuses on the crucial text-based features based on the attended visual features, followed by the self-attention unit to extract the sentiment-rich multimodal features.

- To show our method's effectiveness, we conduct experiments on four real-world datasets: 1) Strongly labelled: MVSA-Single and MVSA-Multiple datasets, 2) Weakly labelled: Flickr and Getty images datasets. The performance of our model is validated in terms of Accuracy, Precision, Recall, F1 score, ROC curves, and PRC curves metrics, which confirm the superiority of our model.
- Finally, ablation experiments and visualizations of attention maps are done to analyze the impact of attention mechanisms on the visual and textual data samples for multimodal sentiment classification.

The rest of the manuscript is organized as follows: Section 2 discusses the related work in multimodal sentiment analysis and multimodal fusion. Section 3 elaborates on our proposed network. Section 4 presents the experimental results, and Section 5 concludes our work and outlines future work.

## 2 RELATED WORK

This section discusses the work related to the field of multimodal sentiment analysis and multimodal fusion.

### 2.1 Multimodal Sentiment Analysis

With the rapid popularity of smartphone devices, an enormous amount of data is generated on social media. This makes sentiment analysis on multiple modalities a popular field of research. The early works have majorly focused on feature selection based approaches. Baecchi et al. [11] applied the continuous bag-of-words (CBOW) model for extracting textual information and denoising autoencoder for getting the robust visual features on Twitter short messages. Fang et al. [12] proposed a probabilistic graphical model to capture the correlation between the textual and visual data of Flickr. Ji et al. [10] proposed a hypergraph learning framework which computes the relevance among the textual, visual, and emoticon modalities on Sina Weibo microblogs. Dai et al. [13] constructed a structured forest to generate the bag of affective words, which reduces the gap between the low-level features and affective descriptors on Multilingual Visual Sentiment Ontology dataset.

Due to the powerful performance of deep learning-based approaches [14], these techniques are increasingly being applied for multimodal sentiment analysis. Xu et al. [15] applied word-level and sentence-level attention for modeling the textual data and the CNN-LSTM approach for extracting the semantic information in images. Chen et al. [16] used emoticons as weak labels and leaned joint features from the image and textual modalities using CNN and dynamic CNN. A probabilistic graphical model was applied to infer the correlation among the predicted labels of various modalities. Zhao et al. [17] experimented on five pre-trained CNN models for extracting the features from images, and word2vec was applied for textual feature extraction. Cosine similarity was applied to quantity the consistency between the features from both the modalities. Finally, the features were merged for the classification. Yu et al. [18] proposed a network for entity-level multimodal sentiment classification. They extracted and represented the target entity using the LSTM network, followed by capturing the contextual information using the attention mechanism. Bilinear pooling was utilized to capture the interactions among the different modalities.

### 2.2 Multimodal Fusion

Multimodal fusion [19] integrates the features from multiple data sources to predict the final class value. There are mainly three types of fusion strategies: early fusion, late fusion, and intermediate fusion.

Early fusion combines data from the input features of multiple modalities to obtain a single feature vector. Poria et al. [7] extracted the visual and textual multimodal features through deep based networks and fused them using multiple kernel classifier for sentiment classification. However, early fusion cannot capture the time-synchronicity of different modalities and often results in a high dimensional redundant feature vector. Late fusion refers to a combination of results from multiple classifiers, where each classifier is trained on a separate modality. However, late fusion ignores the low-level interaction of the modalities. Xu et al. [8] developed a bi-directional attention model, which exploits the correlation between the visual and textual contents simultaneously to fuse the attended visual-textual features via late fusion. In Xu et al. [20], the cross-modal relation among the images, text, and social links was explored through multi-level LSTMs. A joint relationship was obtained to learn the inter-modal correlations at different levels.

In deep learning models, intermediate fusion is commonly employed as a deep multimodal fusion strategy, where the input data is changed into a higher-level representation through multiple layers. Huang et al. [9] proposed multimodal attentive fusion, which focuses separately on the visual attention model and semantic attention model, followed by intermediate-fusion-based multimodal attention. In intermediate fusion, the different representations are fused at different depths, leading to overfitting where the network fails to model the relationship between each modality. Hence, a careful design approach needs to be followed.

Although the existing works for multimodal sentiment analysis have shown significant improvements, several factors need to be incorporated for effective results. Firstly, current works fail to extract important sentiment words corresponding to the image features. Secondly, they do not utilize the channel dimension to generate robust visual features, which enhances the crucial channels in the given image and boosts the model's overall performance. Thus, we aim to develop a Deep Multi-level Attentive network (DMLANet), which tackles the above challenges.

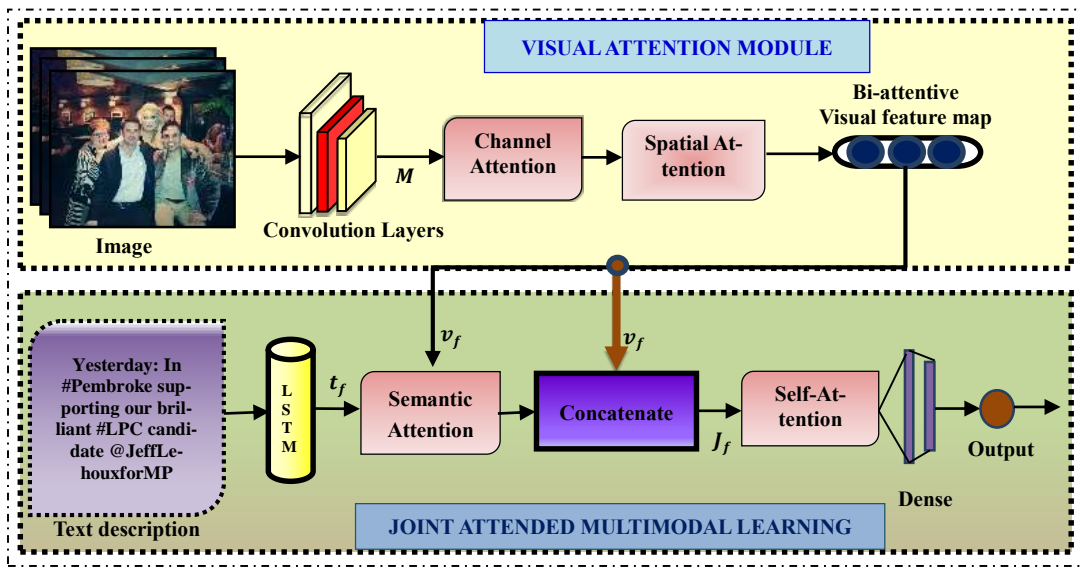

**Fig. 1. Block diagram of the proposed DMLANet**

## 3 PROPOSED METHODOLOGY

This section presents the details of the proposed Deep Multi-level Attentive network (DMLANet). In Section 3.1, we give an outline of the proposed network. Then we present the visual attention module in Section 3.2, which generates significant bi-attentive visual features by utilizing channel attention and spatial attention. Finally, Section 3.3 discusses a joint attended multimodal learning process that learns a combined representation for textual and visual features by applying semantic attention, which measures the semantic closeness of text and visual features, followed by a self-attention mechanism which extracts the crucial multimodal features for sentiment classification.

### 3.1 Framework Overview

Let D represent the given set of documents. For each document $d \in D$, let $I = \{I_1, I_2, \ldots, I_n\}$ denote the set of images for the visual component of the document and $T = \{T_1, T_2, \ldots, T_n\}$ denote the set of text descriptions or the sequence of sentences for the textual component of the document. Each of the sentence $T_i$ is composed of a sequence of $w_i, i \in [1, S]$. Each document is further associated with one of the following sentiment labels: positive, negative, and neutral (Flickr and Getty datasets are labeled with positive and negative sentiments only). Thus, the objective is to predict the sentiment labels on the unseen documents by training the network on the training corpus.

Fig. 1 shows the block diagram of the proposed framework. In the visual attention module, we employ channel-based attention, which enhances the information-rich channels, and spatial or region-based attention, which further concentrates on the emotional regions based on attended channels to get the bi-attentive visual feature map. In joint attended multimodal learning, semantic attention is applied to measure the emotional words related to the bi-attentive visual features. Next, we combine the attended word features and bi-attentive visual features and pass them to the self-attention block, which automatically highlights the important multimodal features. These features are then passed to the classifier for the sentiment classification.

### 3.2 Visual Attention Module

Recently, attention networks have shown significant performance in many computer vision tasks [21]. They make CNN learn and focus on the crucial information by suppressing unnecessary information, thus improving the overall classification performance. We achieve this by sequentially generating the bi-attentive map along the spatial and channel dimensions separately to magnify CNN's representation power. This approach was popularly used for the task of object detection [22]. However, in multimodal sentiment classification, most of the previous works have ignored the channel dimension for obtaining the visual features. This is important because channel-based attention concentrates on the information-rich channels, i.e., they highlight 'what' are the crucial elements in the given image. The upcoming Section 3.2.1, 3.2.2, and Fig. 2 describes the entire visual attention module.

#### 3.2.1 Channel Attention

For each image $I_i$, we obtain the feature map $M \in [H * W * C]$ using the Inception V3 [23] network. In channel attention, we apply global average pooling, and max average pooling of feature maps to generate the average-pooled and max-pooled features, respectively. Each feature is passed to a multilayer perceptron network with one hidden layer followed by the ReLu activation function, and the elements are concatenated to get the final attention map $A_c = (1 * 1 * 256)$. The reason to apply ReLu over tanh is that it converges quickly and results in cheaper computation. The channel attention process can be summarized in Eq (1) as follows:

$$A_c = ReLu\,[W_1\left(W_0\left(MP\,(M)\right)\right) + W_1\left(W_0\left(GAP\,(M)\right)\right)] \quad (1)$$

where, $W_0, W_1$ are the weights of the multilayer perceptron, $M$ denotes the feature map, $MP$ = max-pooling layer, $GAP$ = global average pooling layer, and $A_c$ is the channel attention. Hence, channel attention extracts '*what*' are the meaningful features in a given image by squeezing the spatial dimensions of the feature map using the average and max-pooling layers and merging the output vectors using element-wise summation.

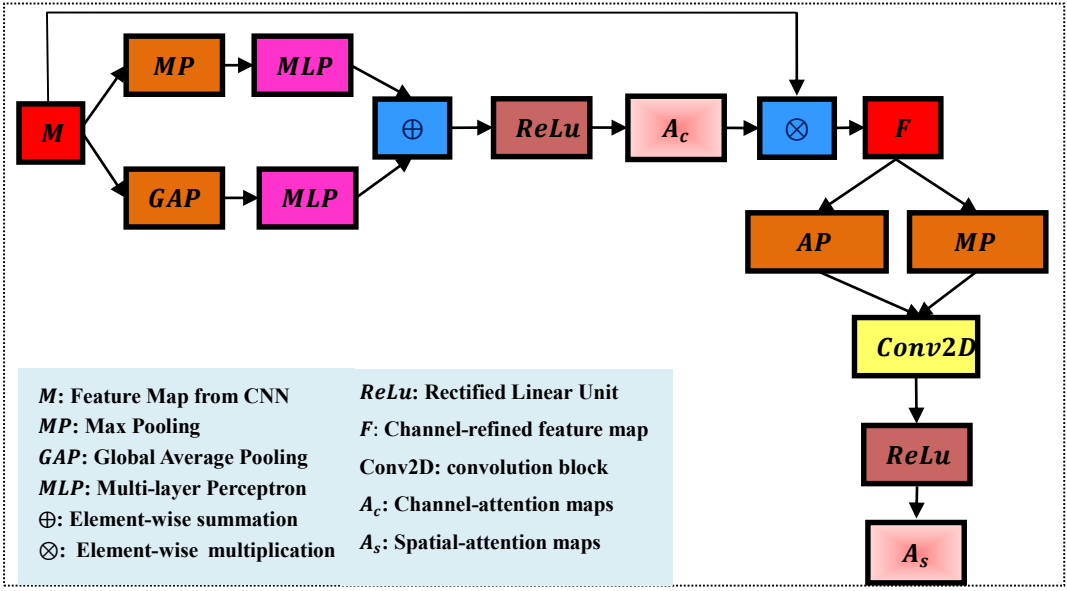

**Fig. 2 Block diagram explaining the Visual Attention Module**

### 3.2.2 Spatial Attention

The spatial attention map tells '*where*' is the informative part of the image, i.e., it locates the relevant image regions according to the attended channel based features. The input feature map $M$ is element-wise multiplied with the channel attention map $A_c$ to generate the channel refined feature map $F$. The channel refined feature maps are concatenated using average pooled and max pooled layers and are fed into a convolutional layer with 7*7 kernel size to generate the spatial attended features, which we refer to as the bi-attentive visual features as they are the combination of the channel-attended and spatial-attended visual features. The process of converting the channel refined feature map $F$ into the spatial attended features $A_s$ can be summarized in Eq (2) as follows:

$$A_s = ReLu \left[ Conv2D \left( AP(F); MP(F) \right) \right] \quad (2)$$

Finally, we obtain the following sequence of bi-attentive visual features, as shown in Eq (3) below:

$$v_f = \{v_1, v_2, \ldots, v_n\}, v_f \in R^{m*d} \quad (3)$$

Where, m= number of regions, and d = feature dimension of each region.

### 3.2.3 Joint Attended Multimodal learning

Each word $w_i$ is transformed into a real-valued vector through pre-trained embedding matrix (we use Glove[1] embeddings), and then optimized by applying LSTM (Long-Short Term Memory) network which gives the high-level textual features as $t_f$. Existing work fails to detect the sentimental words that are related to the images. However, we address this problem by exploiting the correlation between the words and the different visual features generated in Section 3.2. We measure the semantic closeness of the textual features with visual content by combining both the features using element-wise multiplication and generating the joint features $m_f$. This is shown in Eq (4) as follows:

$$m_f = \tan h \left( W(v_f \odot t_f) \right) \quad (4)$$

Where, W are the learnable weights.

The attention scores are computed, as shown in Eq (5) below:

$$\alpha_f = \frac{\exp(m_f)}{\sum_f \exp(m_f)} \quad (5)$$

Finally, we obtain the attended word-level features, which measures the emotional textual features related to the visual features as follows:

$$s_f = \sum_f \alpha_f * t_f \quad (6)$$

Next, we concatenate the obtained attended textual features $s_f$ with the visual features to obtain the joint-multimodal features $J_f = (s_f, v_f)$. Since, in multimodal learning, not all the modalities contribute equally in the classification task [24]. Hence, we apply self-attention networks, which take the multimodal feature vectors as input and automatically identifies the crucial weights corresponding to each modality, as shown in Eq (7) below:

$$v_f = \frac{\exp(\varphi(W*J_f + b))}{\sum_f \exp(\varphi(W*J_f + b))} \quad (7)$$

Where, W and b are the learnable weights, and $\varphi$ is the activation function.

Thus, in a self-attention network, multiple input modalities are allowed to interact with each other to find the input that gets more attention, which tells the importance of the different multimodal input features in the sequence. The joint attended multimodal features are computed as the weighted average over all the feature sequence, as shown below:

$$M = \sum_f v_f * J_f \quad (8)$$

---

[1] https://nlp.stanford.edu/projects/glove/

The obtained attended multimodal features $M$ are passed as an input to the softmax classifier for sentiment classification as follows:

$$P(s) = Softmax(W_s; M) \quad (9)$$

The whole network is trained on a training set by minimizing the cross-entropy loss with backpropagation as follows:

$$Loss = -\sum \log(P(s), y) \quad (10)$$

Where y is the actual sentiment label of the training data.

## 4 EXPERIMENTS

In this section, we conduct several experiments to confirm the efficacy of our DMLANet on popular real-world datasets and report the quantitative and qualitative results.

### 4.1 Datasets

We collected four large-scale, real-world datasets from various social media platforms for conducting the multimodal sentiment classification. Table I shows the complete statistics of each dataset. Further, the datasets are explained as follows:

#### A. MVSA (Multi-View Sentiment Analysis Dataset):

The MVSA dataset [25] consists of two separate datasets. MVSA-Single which contains 5129 image-text pairs from Twitter, where each pair is labelled by a single annotator. MVSA-Multiple consists of 19600 image-text pairs which are labelled by three annotators. The actual sentiment is calculated by taking the majority vote out of the three sentiments (positive, negative, and neutral) for each modality separately. In both cases, the annotator's judgment for the text and image sentiment label is independent. However, many tweets may result in inconsistent textual and image sentiment label. We follow the following rules to deal with inconsistent sentiment labels between different modalities: The tweets with one positive label and one negative label or vice-versa are removed. If the tweet has one positive (or negative) label and other neutral label, then the final multimodal sentiment is positive (or negative). Finally, we get 4511 image-text pairs for MVSA-Single and 17024 image-text pairs for MVSA-Multiple datasets, respectively.

#### B. Flickr:

We collect the image-text pairs from the Flickr website by using the 1200 adjective-noun pair (ANPs), as described in [26]. The images were weakly labeled according to the sentiment of the ANP into the positive and negative sentiment category only. We also collect the English descriptions associated with the images. The images with too short text (<5 words) and too long text (>100 words) were removed. Thus, we obtained a dataset of 276,571 weakly labelled image-text pairs.

#### C. Getty Images:

Getty Images is a supplier of videos, photos, music having relatively formal text descriptions, which can be conveniently browsed by the users. Similar to [27], we query Getty images with 101 sentimental keywords from the Balanced Affective Word List Project[2] to download the image with their corresponding text description. The downloaded image-text pairs were weakly labeled as per the sentiment keywords into the positive and negative sentiment category, giving us 453,289 image-text pairs.

Table I
OVERALL STATISTICS OF EACH DATASET

| Datasets | #Positive | #Negative | #Neutral | Total | Label |
|---|---|---|---|---|---|
| MVSA-Single | 2683 | 1358 | 470 | 4511 | Strong |
| MVSA-Multiple | 11318 | 1298 | 4408 | 17024 | Strong |
| Flickr | 129317 | 147254 | - | 276,571 | Weak |
| Getty Images | 235732 | 217557 | - | 453,289 | Weak |

### 4.2 Implementation Details

The proposed DMLANet is implemented in Python using Keras deep learning framework. The experiments were performed on a 64-bit Windows 10 machine with 128 GB RAM and NVIDIA Titan-RTX GPUs. We set the learning rate = 0.001, batch size = 256 with Adam optimizer. Dropout is used to avoid overfitting. We performed experiments via a five-fold cross-validation strategy. The datasets are split in the 80:10:10 ratio for training, validation, and testing sets, respectively. The final accuracy is calculated by averaging the results across each of the test fold. The model achieving the highest validation accuracy is selected for the testing phase.

### 4.3 Results and Analysis

In this section, we validate the proposed model on all four datasets, as shown in Fig. 3. We use the following evaluation metrics: Precision, Recall, F1 score, and accuracy to validate our model. All the evaluation metrics are ranged from 0 to 100%, where higher the value of the metrics, the better is the performance of the model. For multi-class classification, the average F1 score and accuracy are 79.59% and 79.47%, respectively for MVSA-Single, and 75.26%, and 77.89%, respectively, for MVSA-Multiple. For binary-class classification, the average F1 score and accuracy are 89.19% and 89.30%, respectively for Flickr, 92.60%, and 92.65% respectively for Getty images.

Since, we have imbalanced samples in the dataset, hence we used ROC (Receiver operating characteristics) and PRC (Precision-Recall curves) to validate the performance of our model further. The ROC curves in Fig. 4 (a) shows that our model has shown increased TPR (True Positive rates) on all the datasets. Similarly, AUC (Area under ROC curves) helps to compare the different ROC curves better. The highest value of AUC is 94.46%, which is achieved by Getty images. However, still the model can distinguish between the classes for both the binary and multi-class sentiment classification. Compared with ROC, the PRC are more suitable for imbalanced datasets. Hence, we plotted the PRC in Fig. 4 (b) between the precision and recall values to compare the performance of our model across the datasets. As evident from the curves, the joint attended learning approach in DMLANet has shown effective results for learning the multimodal features for the sentiment classification.

---
[2] http://www.sci.sdsu.edu/CAL/wordlist/origwordlist.html.

and Getty Images datasets.

### 4.4.1 MVSA Datasets:

For MVSA-Single and MVSA-Multiple datasets, following

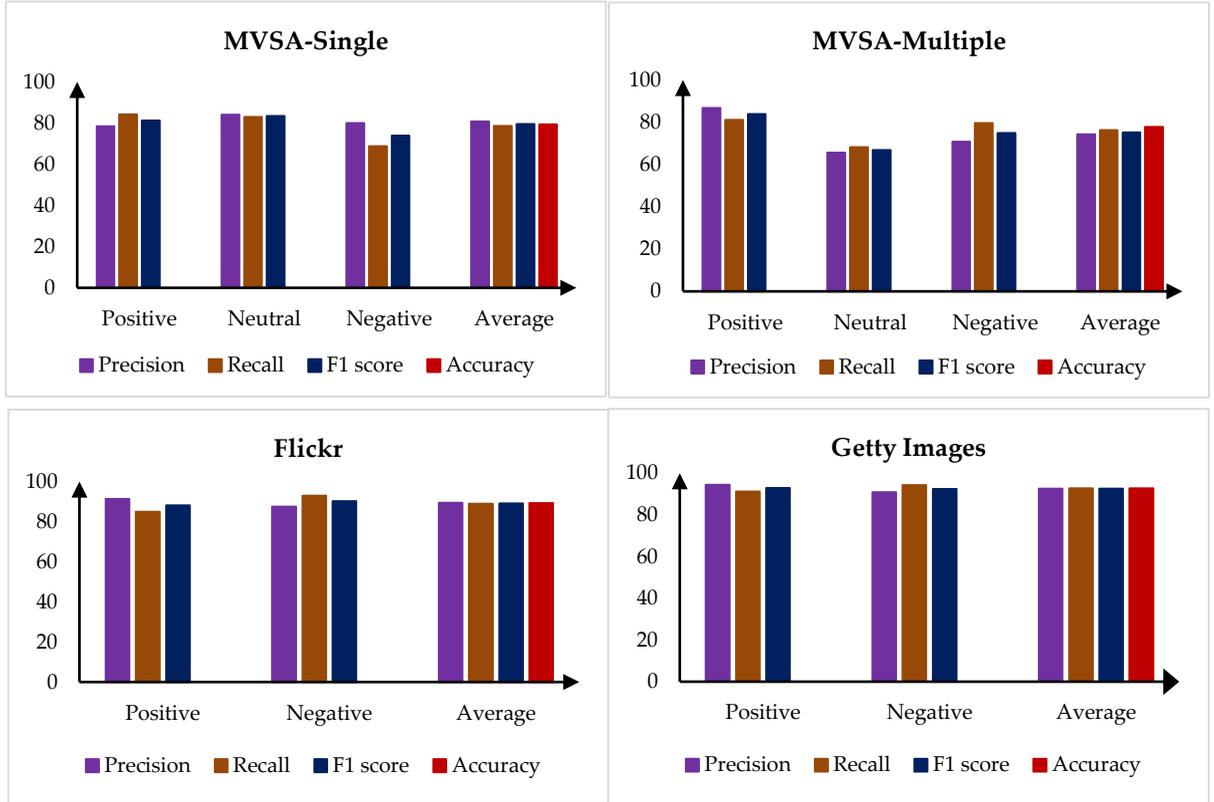

**Fig. 3 Experimental results on the datasets (%)**

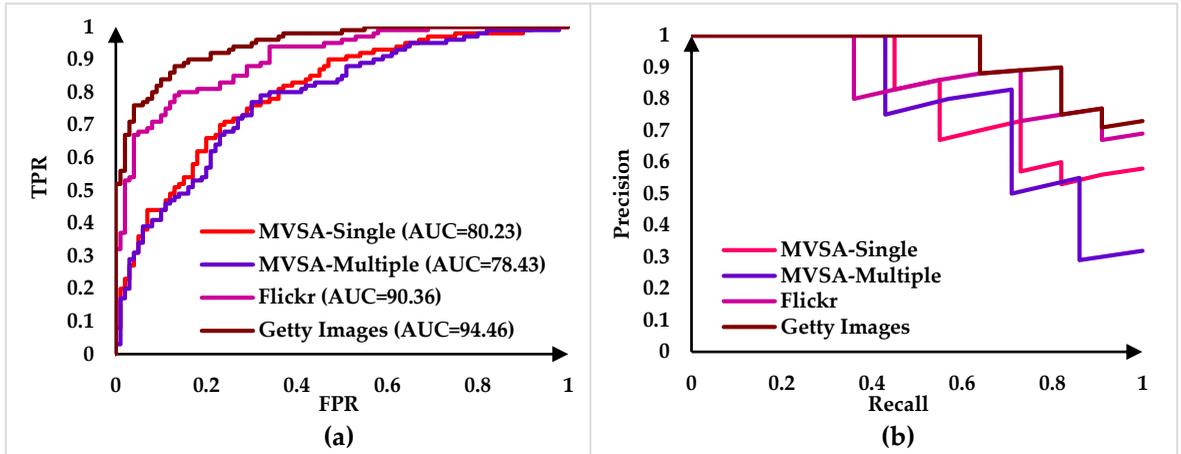

**Fig. 4 (a) ROC curves (b) PRC curves for the datasets**

To show the evolution of model's performance, we plot the training and validation curves for MVSA-Multiple dataset. Fig. 5 (a) shows that training and validation loss decreases with increasing data and (b) shows that training and validation accuracy increases with the data. It can be clearly seen that as more and more data is supplied to the model, it can learn the adequate features from the data and finally converges after approximately 50 epochs.

### 4.4 Baseline Methods

This section compares our work with state-of-the-art methods for the MVSA-Single, MVSA-Multiple, Flickr, baselines were used for comparison:

- **SentiBank and SentiStrength** [26]**:** The SentiBank extracts 1200 ANP as mid-level features for image classification, and SentiStrength utilizes grammar and spellings style from the text. Both the techniques are combined to handle the multimodal sentiment classification.
- **MNN (Merged Neural Network)** [28]**:** MNN utilizes CNN to extract the multimodal features which are fused by residual model using early (Early-RMNN) and late fusion (Late-RMNN).
- **HSAN (Hierarchical Semantic Attentional Network)**

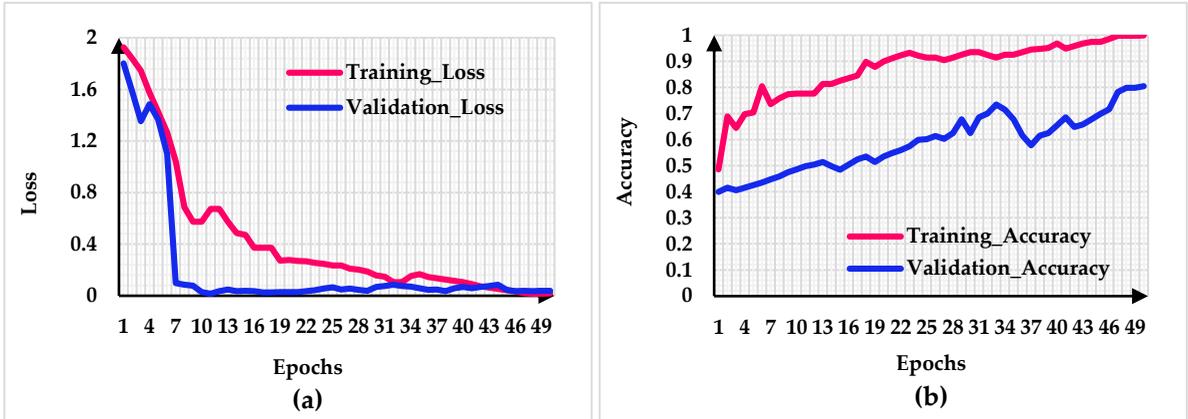

Fig. 5 (a) Training and Validation Loss curves (b) Training and Validation accuracy curves on MVSA-Multiple Dataset

[15]: The text-based HAN extracts textual information from tweets, and semantic image features are extracted by CNN-LSTM model. The important words are reflected by using the attention mechanism.

- **CoMN (Co-Memory Network)** [29]**:** A stacked co-memory network is developed, which uses text features to capture image feature maps and image information is utilized for identifying the textual keywords.
- **MultiSentiNet** [30]**:** This model extracts the objects and scenes from the image, followed by attention-based LSTM to fetch the textual features. Finally, the features are fused for the final sentiment classification.
- **FENet (Fusion-Extraction Network)** [34]**:** It uses Interactive Information Fusion (IIF) mechanism, which applies attention features across both the modalities and Specific Information Extraction layer (SIE), which is based on gated convolution, followed by late fusion for combining both the modalities.

**Table II**
**COMPARISON RESULTS OF DIFFERENT METHODS FOR MVSA DATASETS (%)**

| Methods | Datasets | MVSA-Single | | MVSA-Multiple | |
|---|---|---|---|---|---|
| | | F1 | Accuracy | F1 | Accuracy |
| [26] | SentiBank & SentiStrength | 50.08 | 52.05 | 55.36 | 65.62 |
| [28] | Late-RMNN | - | 67.09 | - | 67.90 |
| [15] | HSAN | - | 66.90 | - | 67.76 |
| [29] | CoMN | 70.01 | 70.51 | 68.83 | 68.92 |
| [30] | MultiSentiNet | 69.63 | 69.84 | 68.11 | 68.86 |
| [34] | FENet | 74.06 | 74.21 | 71.21 | 71.46 |
| **Ours** | **DMLANet** | **79.59** | **79.47** | **75.26** | **77.89** |

Table II displays the comparative results for MVSA-Single and MVSA-Multiple datasets. As compared to the best performing baseline, FENet [34], our model achieves 5% more accuracy for MVSA-Single and 6% more accuracy for MVSA-Multiple dataset. This shows that our multi-level attention contributes to the fine-grained features for sentiment classification. Hence, it is evident that the F1 and accuracy scores of our DMLANet are higher than all the other baselines.

### 4.4.2 Flickr and Getty images Datasets:
For Flickr and Getty images, we compare our work with the following baselines:
- **AHRM (Attention-based Heterogeneous Relational Model)** [31]**::** The visual features are captured by dual attention mechanism, followed by graph convolutional network which combines the social context information.
- **BDMLA (Bi-directional Multi-level Attention)** [8]: The joint learning is done by two independent networks that learn the visual attention and semantic attention, followed by fusing the modalities with MLP.
- **AMGN (Attention-based modality Gated networks)** [32]: It utilizes visual and semantic attention model to obtain word-related visual features, followed by gated LSTM to extract more emotional features in the visual and textual modalities.
- **HDF (Hierarchical Deep Fusion)** [20]: HDF captures the correlations between the image and textual content by using hierarchical LSTM. Late fusion is employed using MLP.
- **DMAF (Deep Multimodal Attentive Fusion)** [9]: DMAF uses deep CNN to extract the visual features and LSTM based semantic attention for modelling the textual data.
- **Joint Cross-modal model** [33]: The textual features are computed using attention-based GRU, and visual features are calculated using maximum mean discrepancy. Finally, attention-based LSTM is used to compute the final-sentiment polarity.

**Table III**
**COMPARISON RESULTS OF DIFFERENT METHODS FOR FLICKR AND GETTY IMAGES (%) (%)**

| Methods | Datasets | Flickr | | Getty images | |
|---|---|---|---|---|---|
| | | F1 | Accuracy | F1 | Accuracy |
| [31] | AHRM | 87.5 | 87.1 | 88.4 | 87.8 |
| [8] | BDMLA | 84.8 | 84.9 | 86.2 | 86.5 |
| [32] | AMGN | 86.8 | 87.3 | 88.7 | 88.2 |
| [20] | HDF | 86.1 | 85.9 | 88.0 | 88.1 |
| [9] | DMAF | 85.0 | 85.9 | 86.6 | 86.9 |
| [33] | Joint Cross-modal model | - | - | 81.0 | 80.6 |
| **Ours** | **DMLANet** | **89.19** | **89.30** | **92.60** | **92.65** |

The comparative results on Flickr and Getty images are shown in Table III. The accuracy obtained on Flickr dataset is 89.30% and on Getty images is 92.65%, which is 4% higher than AMGN [9], which performs best amongst the baselines. We also see that the accuracy and F1 scores on Getty images are higher than Flickr. This may be because, as compared to Flickr, the textual descriptions on Getty are more formal and relevant to the image content. Thus, we can say that our proposed model effectively exploits the correlation between the textual and image modalities for all four datasets.

### 4.5 Ablation Study

In this section, we perform an ablation study to evaluate the contribution of each module. We conduct an ablation study on two datasets: MVSA-Multiple (Multiclass and Strongly labelled dataset) and Flickr (Binary class and Weakly labelled dataset). We retrain our model by ablating the following crucial components: Spatial attention (SA) + channel attention (CA), Semantic attention (SMAtt) and Self-Attention (SAtt). The results are shown in Table IV.

**TABLE IV**
**ABLATION STUDIES ON MVSA-MULTIPLE AND FLICKR DATASETS**

| Datasets | Model | F1 score (%) | Accuracy (%) |
|---|---|---|---|
| MVSA-Multiple | DMLANet w/o (SA + CA) | 71.29 | 70.85 |
| | DMLANet w/o (SMAtt) | 70.17 | 70.00 |
| | DMLANet w/o SAtt | 73.98 | 73.54 |
| | **DMLANet** | **75.26** | **77.89** |
| Flickr | DMLANet w/o (SA + CA) | 85.54 | 85.77 |
| | DMLANet w/o (SMAtt) | 82.44 | 81.90 |
| | DMLANet w/o SAtt | 88.01 | 87.98 |
| | **DMLANet** | **89.19** | **89.30** |

- **DMLANet w/o (SA + CA):** This ablated model gives a drop in the F1 score and accuracy values for both the datasets, which clearly shows the channel-attended visual features and region-attended features helps in learning the discriminative image features.
- **DMLANet w/o (SMAtt):** Here, we ablate the semantic attention block and directly concatenate the bi-attentive visual features with the high-level textual features obtained from LSTM. In this case, we observe a significant drop in the performance of the model for both the datasets. Around 7% accuracy is dropped for MVSA-Multiple dataset, and 8% is dropped for the Flickr dataset. These results indicate the importance of semantic attention, which tells how closely the words are linked to the contents of the images. Thus, it explores the correlation between both the features of the modalities.
- **DMLANet w/o SAtt:** In this ablated model, the self-attention module is not used. The joint multimodal features $J_f$ are directly fed into the dense layer for the final classification. We observe that the F1 score drops by 2% and 1% for MVSA-Multiple and Flickr datasets, respectively. Similarly, the accuracy drops to 73.54% and 87.98% for both the datasets. These results also show that it is necessary to focus only on the essential sentiment-rich multimodal features, as not all the features are important for the classifier.

Based on the results in Table IV, we conclude that the doesn't use the spatial and channel attention block. The features obtained from the inception V3 module are directly given to the semantic attention block. This multi-level attention in the form of channel attention, spatial attention, semantic-attention, and self-attention exploits the correlation between the visual and textual modalities by filtering out the irrelevant and redundant information. This enhances the performance of the multimodal data for sentiment analysis.

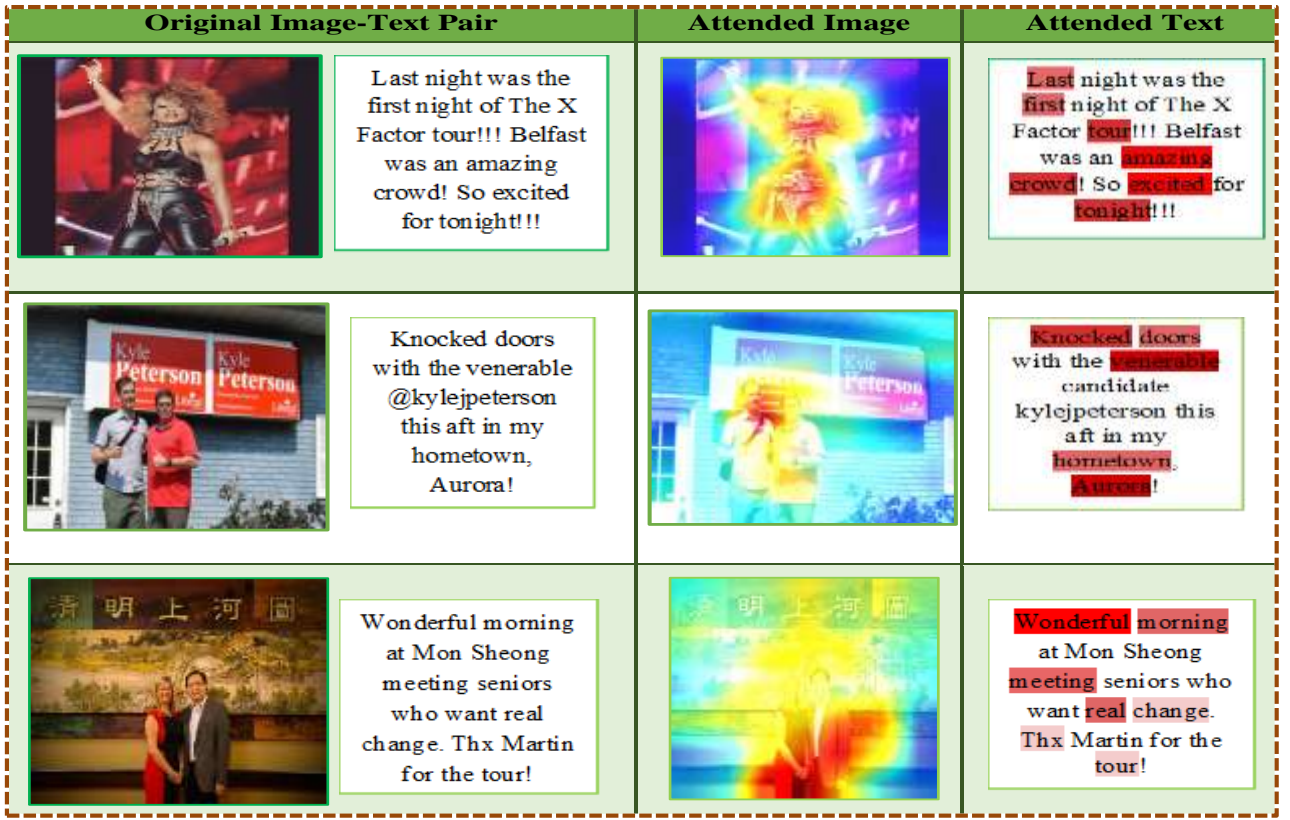

**Fig. 6 Quantitative analysis of DMLANet for Positive image-text pairs**

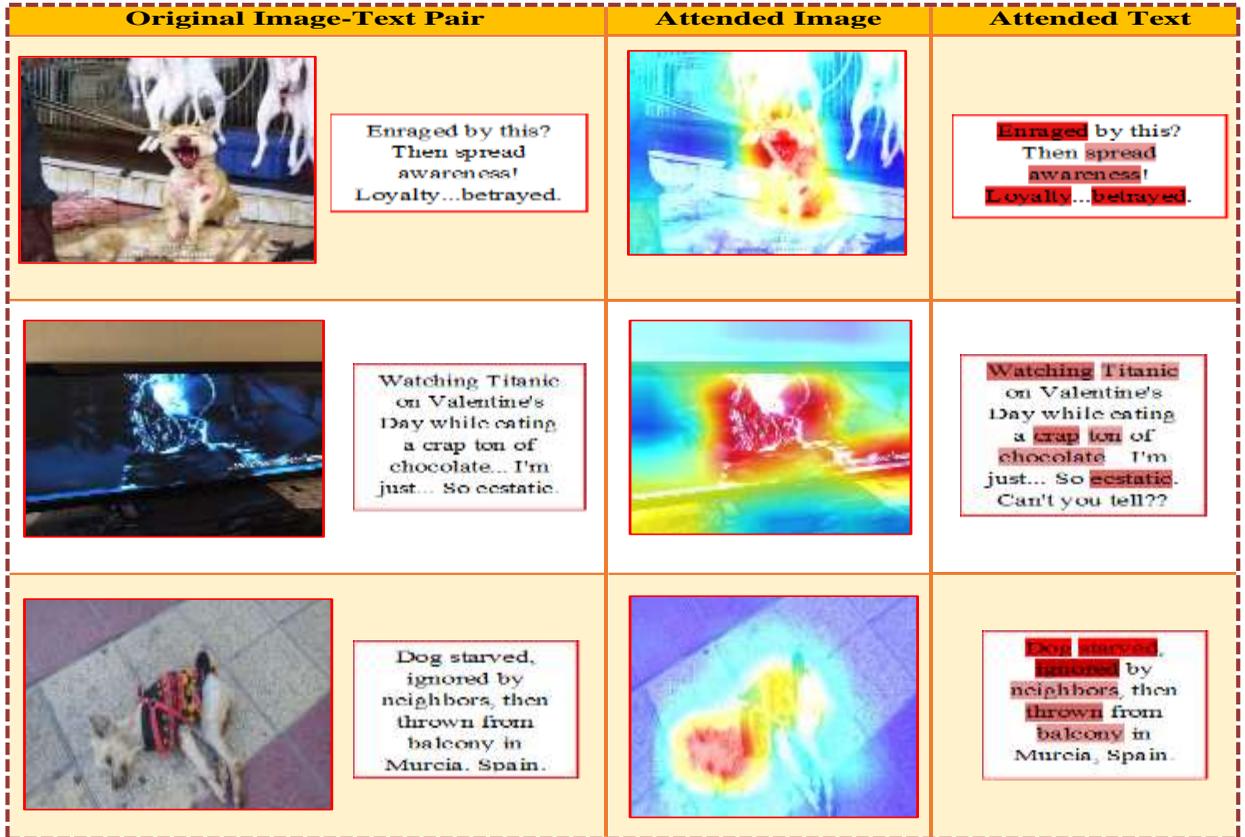

**Fig. 7 Quantitative analysis of DMLANet for Negative image-text pairs**

### 4.6 Visualization

In this part, we evaluate our proposed model by quantitatively showing the sentiment classification results. We randomly select three positive samples (Fig. 6) and three negative samples (Fig. 7) from the MVSA datasets. We use gradient-based class activation maps [35] to visualize the visual attention weights, whereas the background color reflects the semantic attention. The brighter the color, the higher is the attended semantic score. Together, the visual and semantic attention tells "what" our model infers from the image and text sentiment pair.

As seen in Fig. 6, the visual attention is drawn from right image regions by paying attention to more affective regions, which contributes towards the positive sentiment. The semantic attention focuses on words like "amazing", "excited", "wonderful", which conveys the positive sentiment. Similarly, in Fig. 7, negative sentiments are expressed by focusing on crucial regions and words like "starved", "crap", "betrayed". However, it is difficult to tell the exact text's sentiment in many cases, since a text may contain many sarcastic statements where positive words may sarcastically convey negative sentiments. For e.g., In Fig 7, the second example uses some positive words like "ecstatic", still it conveys a negative sentiment. However, combining visual attention helps to classify the sample as negative.

### 5. CONCLUSION

This paper proposes a Deep Multi-level Attentive network (DMLANet) to model the correlation between image and text modalities and extracting only the sentiment-rich multimodal features. The visual attention block applies channel and spatial attention to generate robust bi-attentive visual features. Moreover, the joint-attended multi-modal learning aims to acquire high-quality representations from text and image modalities by focusing on the words that are related to the image contents. Experimental results on four real-world datasets show promising results, as validated by Precision, Recall, F1 score, accuracy, ROC, and PRC metrics. Hence, the proposed model achieves the best performance as compared to other multimodal based approaches.

The proposed model focuses on data samples having some fine-grained correlation between image-text pairs. However, this is not true in reality. Hence, we plan to develop a robust fusion method that could work well on datasets that do not have a close cross-modal correlation in the future.

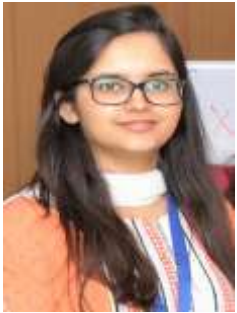

**Ashima Yadav** received B.Sc.(Hons) in Computer Science from University of Delhi, New Delhi, India in 2013, and M.C.A. from Guru Gobind Singh Indraprastha University, New Delhi, India in the year 2016. She is currently working towards the Ph.D. degree from the Department of Information Technology, Delhi Technological University, New Delhi, India. Her current research interest includes deep learning, natural language processing, machine learning, Emotion Recognition, and sentiment analysis.

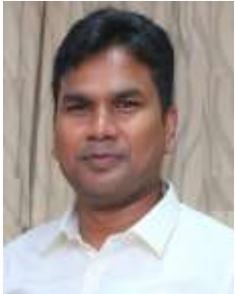

**Dinesh Kumar Vishwakarma (M'16, SM'19)** received the B.Tech. degree from Dr. Ram Manohar Lohia Avadh University, Faizabad, India, in 2002, the M.Tech. degree from the Motilal Nehru National Institute of Technology, Allahabad, India, in 2005, and the Ph.D. degree degree in the field of Computer Vision and Machine Learning from Delhi Technological University University (Formerly Delhi College of Engineering), New Delhi, India, in 2016. He is currently an Associate Professor with the Department of Information Technology, Delhi Technological University, New Delhi. His current research interests include Computer Vision, Machine Learning, Deep Learning, Sentiment Analysis, Fake News and Rumour Analysis, Crowd Behaviour Analysis, Person Re-Identification, Human Action and Activity Recognition. He is a reviewer of various Journals/Transactions of IEEE, Elsevier, and Springer. He has been awarded with "Research Excellence Award" by Delhi Technological University, Delhi, India in 2018, 2019 and 2020.